\documentclass[preprint2]{aastex}

\begin{document}
\title{The PL calibration for Milky Way Cepheids and its implications for the distance scale}

\shorttitle{The PL calibration for Milky Way Cepheids}
\shortauthors{Turner}

\author{David G. Turner} 
\affil{Saint Mary's University, Halifax, Nova Scotia, Canada}
\email{turner@ap.smu.ca}

\begin{abstract}
The rationale behind recent calibrations of the Cepheid PL relation using the Wesenheit formulation is reviewed and reanalyzed, and it is shown that recent conclusions regarding a possible change in slope of the PL relation for short-period and long-period Cepheids are tied to a pathological distribution of {\it HST} calibrators within the instability strip. A recalibration of the period-luminosity relation is obtained using Galactic Cepheids in open clusters and groups, the resulting relationship, described by $\log L/L_{\sun} = 2.415(\pm0.035) + 1.148(\pm0.044) \log P$, exhibiting only the moderate scatter expected from color spread within the instability strip. The relationship is confirmed by Cepheids with {\it HST} parallaxes, although without the need for Lutz-Kelker corrections, and in general by Cepheids with revised {\it Hipparcos} parallaxes, albeit with concerns about the cited precisions of the latter. A Wesenheit formulation of $W_V = -2.259(\pm0.083) - 4.185(\pm0.103) \log P$ for Galactic Cepheids is tested successfully using Cepheids in the inner regions of the galaxy NGC 4258, confirming the independent geometrical distance established for the galaxy from OH masers. Differences between the extinction properties of interstellar and extragalactic dust may yet play an important role in the further calibration of the Cepheid PL relation and its application to the extragalactic distance scale. 
\end{abstract}


\section{Introduction}

A reliable calibration of the Cepheid period-luminosity (PL) relation is of crucial importance for the continued use of classical Cepheids as distance indicators for nearby spiral and irregular galaxies. The well known Key Project of the {\it Hubble Space Telescope} ({\it HST}) to measure the Hubble constant {\it H$_{\it 0}$} to within $\pm10\%$ has been marvellously successful \citep{fr01}, with remaining sources of uncertainty generally consisting of: (i) the importance of metallicity differences between Cepheids, and (ii) the zero-point established for Large Magellanic Cloud (LMC) Cepheids.

Alternate zero-point calibrations for the PL relation are possible for Galactic Cepheids by various techniques, including trigonometric parallaxes, cluster parallaxes, and use of the Baade-Wesselink method. The last two techniques generally confirm the existing PL calibration \citep{tb02,fo07}, while the use of parallax data to test the calibration has traditionally been limited by the small size of trigonometric parallaxes for Cepheids relative to their uncertainties, the situation applying to most {\it Hipparcos} parallaxes \citep{fc97,go00}. The technique has been revived recently by \citet{bn02,bn07} for 10 Galactic Cepheids using the {\it HST} Fine Guidance Sensor 3 and grids of surrounding faint stars. A revision of the {\it Hipparcos} results has also been made by \citet{vl07} for a slightly larger sample, with comparable precision in some cases.

An interesting feature of the {\it HST} calibration of the PL relation is that the slope appears to depend upon a single long-period Cepheid in the sample, namely $\ell$ Car. Although that does not appear to affect the zero-point of the relationship \citep{bn07,vl07,fo07}, it is a characteristic that has a ready explanation in terms of the star's location in the Cepheid instability strip, which includes stars of different effective temperatures and surface gravities. Stars of similar pulsation period may have quite different parameters, with more massive stars lying towards the hot side of the strip and less massive stars towards the cool side of the strip. A Cepheid's exact location in the strip can often be established to a first approximation from an individual Cepheid's light amplitude and intrinsic color, yet there is another diagnostic tool that can be applied using the Cepheid's rate of period change, which is tied closely to mass, the factor dominating the rate at which stars evolve through the instability strip \citep{te06a}. All Galactic Cepheid calibrators with trigonometric parallaxes exhibit well-defined period changes that allow their locations within the instability strip to be specified very accurately. The present paper explores the effect of instability strip location on recent calibrations of the Cepheid PL relation, and presents an alternative version that considers all potential complicating factors.

\section{Reddenings of the Galactic Calibrators}

The use of Cepheids as distance indicators suffers from problems arising from the fact that a Cepheid's luminosity is tied to its location within the instability strip, while its observed brightness is affected by interstellar and/or intergalactic extinction. The situation has been described repeatedly in the literature, and can be summarized through the standard formulation involving the period-luminosity-color (PLC) relation in terms of {\it B--V} color: 
\begin{displaymath}
\langle M_V \rangle = a \log P + b(\langle B \rangle - \langle V \rangle)_0 + c \;,
\end{displaymath}
where {\it a} is the slope of the relationship (the period dependence), {\it b} is the color term, {\it c} is the zero-point, and angled brackets denote intensity means. Reddening affects the observed colors of Cepheids, with intrinsic color being given by:
\begin{displaymath}
(\langle B \rangle - \langle V \rangle )_0 = (\langle B \rangle - \langle V\rangle) - E_{B-V} \;,
\end{displaymath}
in which $E_{B-V}$ is the color excess. The extinction at visual wavelengths is specified by the ratio of total-to-selective extinction, $R = A_V$/E$_{B-V}$, therefore:
\begin{displaymath}
\langle V \rangle_0 = \langle V\rangle - R \:E_{B-V} \;.
\end{displaymath}
Combining the relationships \citep{op88,mf91,ma08} produces an equation for the distance {\it r} to any Cepheid given by:
\begin{eqnarray}
\nonumber
5 \log r = \langle V\rangle - a \log P - b (\langle B \rangle - \langle V \rangle) \\
\nonumber
+ (b-R)E_{B-V}- c + 5 \;.
\end{eqnarray}

The color term {\it b} for the PLC relation is typically close to the value of {\it R} applicable to interstellar extinction, although the exact value is a matter of debate \citep[see][]{op84}. The similarity of the two terms makes it possible to adopt a color correction similar to the value of {\it R} to correct for the effects of either color spread within the instability strip or reddening, but not necessarily both \citep[see][]{mb86,mf91}, unless $b-R= 0$. Reddening-free {\it Q} parameters have been in use for many years \citep[e.g.,][]{vb68b}, although their validity can be questioned given the sizeable range of parameters characterizing interstellar reddening and extinction within our own Galaxy \citep{tu76,tu89,tu96}.

The first attempt to remove the effects of reddening and instability strip color spread from Cepheids is attributed to \citet{vb68a,vb68b,vb75}, who adopted the then-current PLC color term of 2.67 to correct Cepheid magnitudes primarily for dispersion within the strip as well as for most of the effects of interstellar extinction. The alternative Wesenheit function of \citet{ma76,ma82} used a color term of 3.1 to correct primarily for the effects of interstellar extinction, presumably overcorrecting slightly for color spread within the strip. It was recognized that alternative functional dependences would be required for galaxies where different extinction laws applied \citep{vb75,bm80}, and more recent formulations have in fact used a color term of 3.3, a value that is appropriate for nearby Galactic stars with similar colors to Cepheids \citep{mf91}.

\setcounter{table}{0}
\begin{table*}[t]
\caption[]{Parallax Cepheid Reddenings and Instability Strip Parameters.}
\label{tab1}
\centering
\small
\begin{tabular*}{1\textwidth}{lccccccccccc}
\noalign{\smallskip} \noalign{\smallskip} 
\hline \hline \noalign{\smallskip}
Cepheid &$\log P$ & & &E$_{B-V}$ & & &Mean &$(\langle B \rangle$--$\langle V \rangle )_0$ &$\Delta B_{10}$ &$\log \dot P_{10}$ $^7$ &Crossing \\
& &(1) &(2) &(3) &(4) &(5) &(6) & & &(s yr$^{-1}$) \\
\noalign{\smallskip} \hline \noalign{\smallskip}
RT~Aur &0.5715 &0.051 &0.089 &0.050 &0.080 &0.087 &0.071 &0.520 &0.970 &+0.200 &3 \\
T~Vul &0.6469 &0.064 &0.054 &0.068 &0.110 &0.110 &0.081 &0.563 &0.719 &{\it +0.388} &2 \\
FF~Aql &0.6504 &0.224 &0.191 &0.224 &0.250 &0.228 &0.223 &0.531 &0.362 &+0.123 &3 \\
$\delta$~Cep &0.7297 &0.092 &0.087 &0.045 &0.090 &0.094 &0.082 &0.577 &0.939 &{\it --0.201} &2\\
Y~Sgr &0.7614 &0.205 &0.195 &0.182 &0.230 &0.199 &0.202 &0.653 &0.796 &--0.177 &3 \\
X~Sgr &0.8459 &0.197 &0.230 &0.219 & ... &0.241 &0.217 &0.526 &0.675 &+0.275 &3 \\
W~Sgr &0.8805 &0.111 &0.108 &0.079 &... &0.179 &0.119 &0.625 &0.926 &--0.214 &3 \\
$\beta$~Dor &0.9931 &0.044 &0.041 &0.000: &0.070 &0.078 &0.047 &0.756 &0.958 &--0.254 &3 \\
$\zeta$~Gem &1.0065 &0.018 &0.009 &0.031 &0.040 &0.072 &0.034 &0.782 &0.703 &{\it +0.481} &4 \\
$\ell$~Car &1.5509 &0.170 &0.151 &... &... &0.183 &0.168 &1.095 &0.598 &+0.421 &5 \\
\hline \noalign{\smallskip} 
\end{tabular*}
\small{(1) \citet{bn07}; (2) \citet{lc07}, BELRAD; (3) \citet{ko08}; (4) \citet{te87}; (5) \citet{tu01}; (6) adopted value; (7) Italicized values refer to Cepheids undergoing period decreases.}
\end{table*}

Recent best estimates for {\it b} seem to encompass values ranging from 2 to 3 \citep[or $\sim$5--6 according to][]{op84}, so a better approach would at first sight seem to be to use regression techniques on Cepheid samples to establish an empirical correction term \citep[e.g.,][]{st90,ma08}. \citet{mf91} argue otherwise. Recent calibrations typically include such a parameter based upon the Wesenheit formulation with magnitudes from the near-infrared spectral region \citep[e.g.,][]{bn07,vl07,fo07}. Use of near-infrared magnitudes has the beneficial effect of reducing, or possibly eliminating, effects that might be linked to reddening law variations or to metallicity differences between Cepheids, which affect colors (mainly in the optical region) corresponding to specific effective temperatures as well as the degree of penetration into the strip for post-hydrogen-burning stars. But the question remains as to whether or not differences in the properties of interstellar dust from one galaxy to another are important enough to affect the calibration used to establish Cepheid distances.

Given the importance to the PL calibration of reddening and the location of individual Cepheids within the instability strip, it is of interest to consider the other methods of calibration available. The reddening of the Cepheids studied by \citet{bn02,bn07} was established from the colors and spectral types of stars lying along the same lines of sight, i.e. from their space reddenings. In many cases the stars also have independent estimates for their reddenings from spectroscopic methods \citep{ko08} or their equivalents \citep{te87}, and some from independently-derived space reddenings \citep{tu01,lc07}, so it is informative to compare the results.

Table~\ref{tab1} compares available estimates for the reddening of {\it HST} Cepheids, as obtained from space reddenings \citep{tu80a,tu84,bn07}, spectroscopic methods \citep{te87,ko08}, and compilations standardized to space reddenings \citep{tu01,lc07}. Photometric reddenings were omitted from such a comparison because they do not always account for the natural dispersion within the instability strip of Cepheids at constant period \citep[see also][]{mf91}. An attempt was made to deduce independent field reddenings for the Cepheids using 2MASS data \citep{cu03}, as illustrated by \citet{te08}, but relatively large uncertainties in the infrared data, in combination with the small reddenings and distances to the stars, produced results with very large uncertainties, much larger than values typically no larger than $\pm0.02$ that apply to the color excesses cited in Table~\ref{tab1}.

\begin{figure}[t]
\begin{center}
\includegraphics[width=0.45\textwidth]{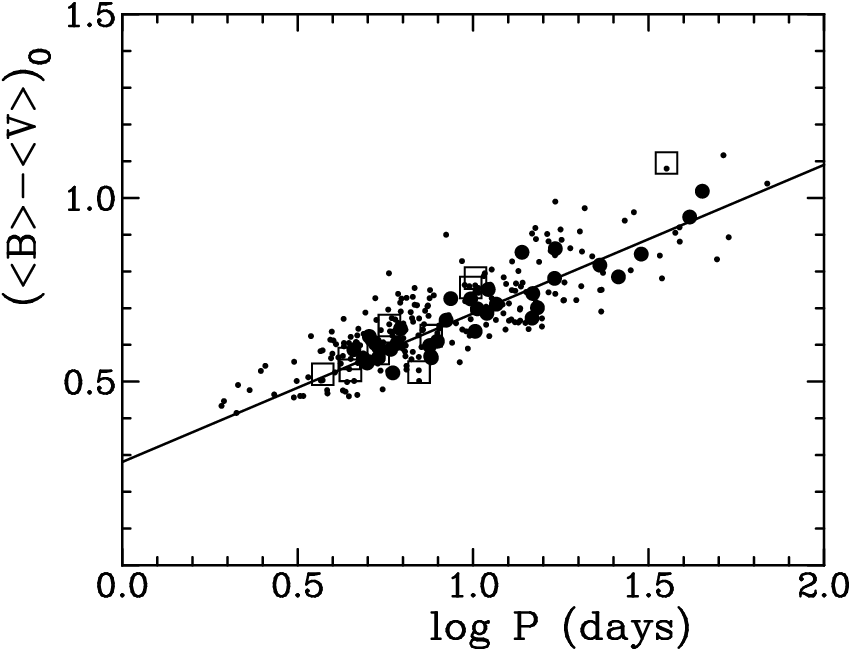}
\end{center}
\caption{\small{The period dependence of intrinsic color for Cepheids of well-established space, spectroscopic, and photometric reddening (small points), Cepheids from the same group with the largest pulsation amplitudes for a given period (large points), and HST Cepheids (open squares). The plotted relationship is a least squares fit to the Cepheids of largest amplitude, and should represent the colors of Cepheids populating the center of the instability strip.}}
\label{fig1}
\end{figure}

The differences between the independent reddening estimates in Table~\ref{tab1} is no larger than the associated uncertainties in the values, so a simple unweighted average was formed for the mean reddening of each Cepheid. Those values are listed in column 8 of Table~\ref{tab1}, and the inferred intrinsic colors are plotted as a function of logarithm of pulsation period in Fig.~\ref{fig1}. Also plotted are intrinsic colors for Cepheids of known reddening, based upon an unpublished compilation of photometric and spectroscopic reddenings tied to a space reddening scale \citep{tu01}, and intrinsic colors for Cepheids in the same sample that have the largest amplitudes for their pulsation period. The latter should concentrate towards the center of the instability strip or slightly blueward, where Cepheids of largest light amplitude are located. The scatter should closely represent the actual scatter caused by the natural dispersion of Cepheids within the instability strip, thus the points corresponding to the {\it HST} Cepheids provide direct information about where they lie in the instability strip.

An inspection of the data leads to the unexpected result that the three Cepheids of longest period in the {\it HST} sample are redder (cooler) than Cepheids lying near the center of the instability strip, while the bluest (hottest) Cepheids in the group are all at the short period end. The three longest period Cepheids in the {\it HST} sample are therefore less luminous than Cepheids lying near strip center, so any attempt to construct a period-absolute magnitude relation based solely on the sample would pathologically underestimate the true slope of the relationship. Just such an effect was noted in the \citet{bn07} study, without the natural explanation evident from the intrinsic colors of the Cepheids. The resulting study of the Cepheid PL relation from {\it HST} parallaxes therefore contains a potential source of bias. The bias in this case is a curious variant of the expected Malmquist bias in that the less-luminous short-period Cepheids in the sample are drawn from the most luminous objects of the class while the more-luminous long-period Cepheids are drawn from the least luminous objects of the class.

\section{Instability Strip Location}

A combination of intrinsic color, light amplitude, and rate of period change can be used to identify the location of each Cepheid calibrator in the instability strip uniquely, as noted by \citet{te06a,te06b}. In essence, a measure of period change for a Cepheid relates directly to the rate at which the progenitor star is evolving through the instability strip. For a specific pulsation period, that rate correlates closely with the mass of the star: more massive stars evolving faster than less massive stars. The correlation is essentially exact for the rapid first crossing of the instability strip \citep{te06a}, but contains more scatter for higher order crossings owing to complications attributable mainly to the degree of envelope contamination of the star by CNO elements \citep{xl04} that are introduced by meridional mixing occurring during previous main-sequence stages \citep{ma01}. The degree to which individual Cepheids penetrate the instability strip following the onset of helium burning and shell helium burning in their interiors is affected by such chemical composition differences, making it difficult, but not impossible, to relate rate of period change to the specific location of individual Cepheids within the instability strip. Light travel time effects in binary systems and stochastic variations in pulsation period are an added complication that affect observed rates of period change in Cepheids, but tend to be only a minor annoyance for the majority of objects. The technique works well for the case of the cluster Cepheid V1726 Cyg \citep{te06b}.

\begin{figure}[t]
\begin{center}
\includegraphics[width=0.45\textwidth]{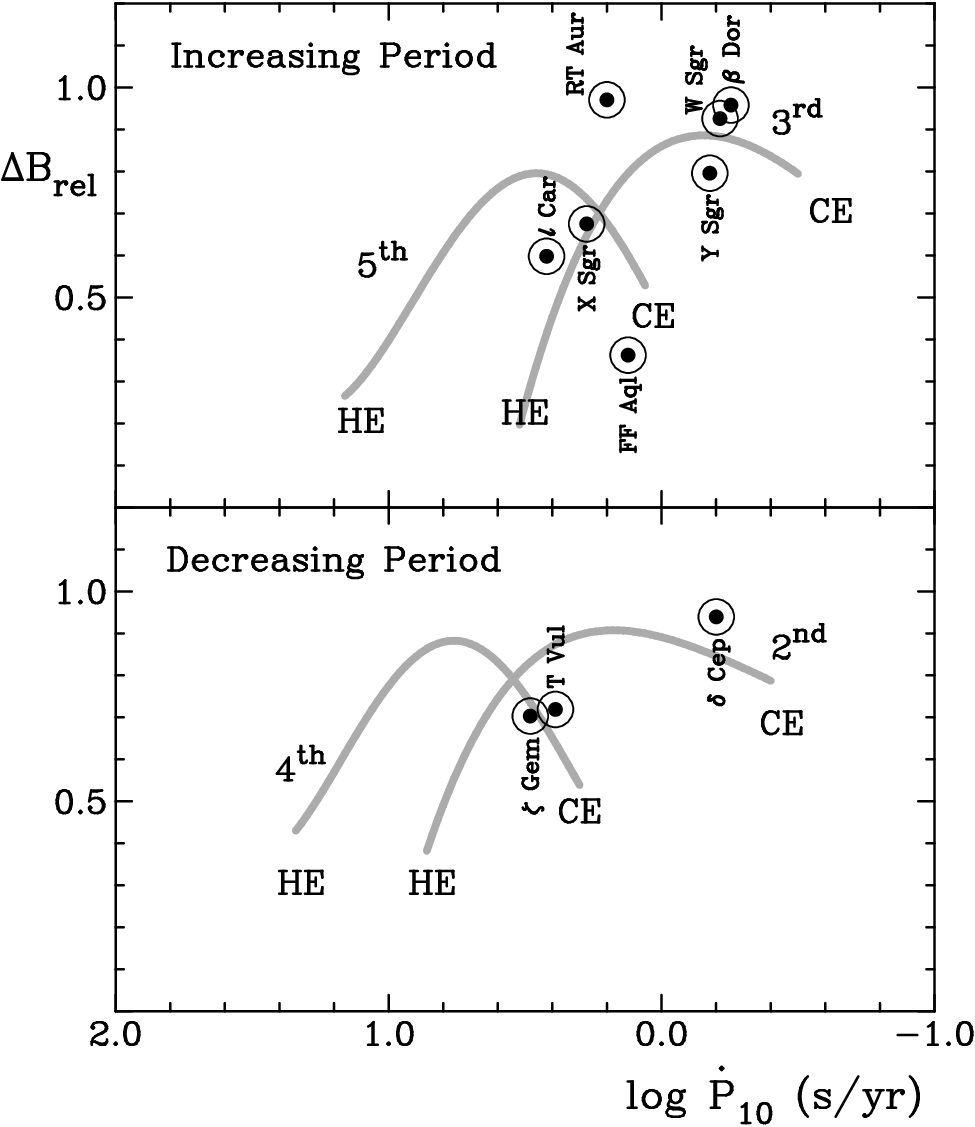}
\end{center}
\caption{\small{The dependence of normalized pulsation amplitude for HST Cepheids (circled dots) relative to normalized rate of period change for Cepheids with well-established period changes (plotted curves). Individual HST Cepheids are identified, as are the hot edge (HE) and cool edge (CE) for Cepheids in third and fifth crossings of the instability strip (upper), and second and fourth crossings (lower).}}
\label{fig2}
\end{figure}

The results obtained for individual {\it HST} Cepheids are described below, with reference to the data of Table~\ref{tab1} as plotted in Fig.~\ref{fig2}. The actual blue light amplitudes and rates of period change for each Cepheid have been normalized to the values appropriate for a Cepheid with a pulsation period of 10 days \citep[see][]{te06a,te06b}, in order to facilitate comparison with the results found for a larger sample of Cepheids. Normalization in such fashion helps to generalize the individual results for a more realistic comparison with the overall results applying to a larger sample of Galactic Cepheids.

\subsection{RT Aur}

RT Aur has a large light amplitude consistent with a location near the center of the instability strip, but with a rate of period increase slightly larger than that of Cepheids of comparable amplitude in the third strip crossing. Its parameters are inconsistent with a fifth crossing of the instability strip, since it would then lie redward of strip center whereas its derived intrinsic color matches those for large-amplitude Cepheids, all of which lie slightly blueward of strip center. The best overall match of its parameters is to Cepheids lying slightly blueward of strip center in the third crossing of the strip. The exact rate of period change for RT Aur is still poorly established because the Cepheid appears to be affected by light travel time effects from its long-term orbital motion about a companion. The rate of period change has only recently been established with confidence \citep{te07}.

\subsection{T Vul}

T Vul has an intermediate light amplitude consistent with a location either redward or blueward of strip  center, and its rate of period decrease matches Cepheids that are either blueward of strip center in the second crossing or redward of strip center in the fourth crossing. Its derived intrinsic color suggests a star slightly hotter than stars of large amplitude lying slightly blueward of strip center. Only a Cepheid in the second crossing of the instability strip lying blueward of strip center has such characteristics.

\subsection{FF Aql}

FF Aql has a small light amplitude, indicating a Cepheid lying well away from strip center, and its reddening places it among stars lying blueward of strip center. Its rate of period increase is reasonably consistent with third crossing Cepheids lying blueward of strip center, but the match is not ideal. Contamination of existing photometry for the Cepheid by its optical companion \citep{ue93} may be an important factor.

\subsection{$\delta$ Cep}

$\delta$ Cep has a reasonably large light amplitude and a rate of period decrease that is consistent with Cepheids in the second strip crossing. Its intrinsic color is also consistent with a star lying near the center of the instability strip in the second crossing.

\subsection{Y Sgr}

Y Sgr has an intermediate-to-large light amplitude and a small rate of period increase, both of which are consistent with a star in the third crossing of the instability strip lying slightly away from strip center. The inferred color indicates a Cepheid lying slightly towards the cool side of the instability strip in the third crossing.

\subsection{X Sgr}

X Sgr has an intermediate light amplitude and a rate of period increase that places it either on the blue edge of the strip for a third crossing or on the red edge of the strip for a fifth crossing. Its very blue color establishes the case for the former quite strongly. X Sgr lies towards the hot edge of the instability strip for a Cepheid in the third crossing.

\subsection{W Sgr}

W Sgr has a large light amplitude and a rate of period increase consistent with Cepheids lying near strip center in the third crossing. Its color is also slightly bluer than those of Cepheids lying near strip center, so, like X Sgr, it must also be a Cepheid in the third crossing of the strip lying blueward of strip center.

\subsection{$\beta$ Dor}

$\beta$ Dor has a large light amplitude and a rate of period increase matching those of Cepheids in the third crossing of the strip lying near strip center. Its color is redder than such stars, suggesting that it lies slightly redward of strip center.

\subsection{$\zeta$ Gem}

$\zeta$ Gem has an intermediate light amplitude and a rate of period decrease that matches Cepheids lying either towards the blue edge of the strip in the second crossing, or towards the red edge of the strip in the fourth crossing. The Cepheid's red color relative to stars lying near strip center is consistent with the latter. Perhaps that attribute has some link to the Cepheid's somewhat unusual light curve relative to other 10-day Cepheids.

\subsection{$\ell$ Car}

$\ell$ Car has a relatively small light amplitude and a rapid rate of period increase, which would normally suggest a Cepheid in the third crossing of the instability strip lying near the blue edge. The star's colors, however, indicate a Cepheid lying well towards the red edge of the strip. The alternate possibility is that $\ell$ Car is in the fifth crossing of the strip lying redward of strip center, which appears to be the only possibility consistent with its observed characteristics. Given its inferred direction of evolution and implied location in the instability strip, $\ell$ Car might be expected to undergo a detectable amplitude decrease in coming years.

\section{Direct Formulation of the PL Relation}

The Wesenheit formulation used for calibration of the Cepheid PL relation has been used so frequently in recent years that it is sometimes overlooked that there are direct means of establishing luminosities for Galactic Cepheids through measurement of their effective temperatures, radii, and bolometric magnitudes. The technique was demonstrated by \citet{tb02}, and involves the use of a very tight period-radius relation \citep[see also][]{ls95,gi98}, a link between intrinsic color and effective temperature as demonstrated by \citet{gr92}, and a bolometric correction scale tied to model stellar atmospheres, in conjunction with measured parameters for the Sun.

\subsection{Cluster Cepheids}

\setcounter{table}{1}
\begin{table*}[t]
\caption[]{Parameters for Cepheid Clusters and Groups.}
\label{tab2}
\centering
\small
\begin{tabular*}{1.00\textwidth}{@{\extracolsep{-1mm}}lrrcclrrc}
\noalign{\smallskip} \hline \hline \noalign{\smallskip}
Cluster &Ref. &$V_0-M_V$ &$E_{B-V}$ & &Cluster &Ref. &$V_0-M_V$ &$E_{B-V}$ \\ 
& & &(Cepheid) & & & & &(Cepheid)\\
\noalign{\smallskip} \hline \noalign{\smallskip}
Cas~R2 &(2) &$7.06\pm0.03$ &$0.27\pm0.02$ & &IC~4725 &(1,10,14,15) &$8.81\pm0.10$ &$0.50\pm0.03$ \\
Turner~9 &(1,3) &$9.33\pm0.05$ &$0.15\pm0.01$ & &NGC~129 &(16) &$11.11\pm0.02$ &$0.51\pm0.01$ \\
ADS~1477 &(4) &$5.02\pm0.07$ &$0.02\pm0.01$ & &NGC~6087 &(17) &$9.78\pm0.03$ &$0.17\pm0.01$ \\
Platais~1 &(5) &$10.98\pm0.02$ &$0.33\pm0.02$ & &ADS~5742 &(18) &$8.00\pm0.10$ &$0.06\pm0.01$ \\
Berkeley~58 &(6) &$12.40\pm0.12$ &$0.64\pm0.02$ & &Lyng\aa~6 &(1,10,14,19) &$11.12\pm0.10$ &$1.36\pm0.05$ \\
NGC~1647 &(7)  &$8.67\pm0.02$ &$0.29\pm0.01$ & &NGC~6067 &(1,10,15,20) &$11.19\pm0.15$ &$0.29\pm0.03$ \\
Dolidze~45 &(1)  &$9.23\pm0.15$ &$0.09\pm0.01$ & &Ruprecht~175 &(21) &$10.43\pm0.04$ &$0.28\pm0.02$ \\
van~den~Bergh 1 &(8) &$11.08\pm0.07$ &$0.75\pm0.02$ & &Trumpler~35 &(1,15,22) &$11.11\pm0.10$ &$0.95\pm0.02$ \\
NGC~5662 &(1,9,10) &$9.28\pm0.05$ &$0.28\pm0.01$ & &Turner~2 &(23) &$11.26\pm0.10$ &$0.56\pm0.01$ \\
Ruprecht~79 &(1,11) &$12.55\pm0.16$ &$0.85\pm0.05$ & &Vel~OB5 &(23) &$11.99\pm0.03$ &$0.35\pm0.02$ \\ 
NGC~6649 &(12) &$11.06\pm0.03$ &$1.27\pm0.02$ & &Vul~OB1 &(1,24) &$11.28\pm0.10$ &$0.59\pm0.03$ \\
Collinder~394 &(1,13) &$9.38\pm0.10$ &$0.30\pm0.02$ & &Anon~Vul~OB &(1) &$12.37\pm0.20$ &$1.02\pm0.03$ \\
\noalign{\smallskip} \hline \noalign{\smallskip} 
\end{tabular*}
\small{(1) This study; (2) \citet{te84}; (3) \citet{te97,te98b}; (4) \citet{tu06}; (5) \citet{te94,te06b}; (6) \citet{te08}; (7) \citet{tu92}; (8) \citet{te98a}; (9) \citet{tu82}; (10) \citet{an07}; (11) \citet{wa87}; (12) \citet{tu81}; (13) \citet{tp85}; (14) \citet{pe85}; (15) \citet{ho03}; (16) \citet{te92}; (17) \citet{tu86}; (18) \citet{tf78}; (19) \citet{wa85a}; (20) \citet{wa85b}; (21) \citet{tu98}; (22) \citet{yi66} and \citet{tu80b}; (23) \citet{te93}; (24) \citet{tu84}.}
\end{table*}

\setcounter{table}{2}
\begin{table*}[t]
\caption[]{Deduced Properties for Cluster Cepheids.}
\label{tab3}
\centering
\small
\begin{tabular*}{0.98\textwidth}{@{\extracolsep{-1mm}}llcccccccc}
\noalign{\smallskip} \hline \hline \noalign{\smallskip}
Cepheid &Cluster &Crossing &$\log P_0$ &$M_V$ &($\langle B \rangle$--$\langle V \rangle$)$_0$ &$\log T_{\rm eff}$ &B.C. &$M_{\rm bol}$ &$\log L/L_{\sun}$ \\
\noalign{\smallskip} \hline \noalign{\smallskip}
SU~Cas &Cas~R2 &3 &0.2899 &$-1.99\pm0.07$ &0.442 &3.807 &--0.01 &--2.00 &2.716 \\
SU~Cyg &Turner~9 &3 &0.5850 &$-2.87\pm0.07$ &0.460 &3.802 &--0.01 &--2.88 &3.069 \\
$\alpha$~UMi &ADS~1477 &1 &0.5987 &$-3.08\pm0.07$ &0.500 &3.793 &--0.01 &--3.09 &3.154 \\
V1726~Cyg &Platais~1 &3 &0.6271 &$-3.06\pm0.07$ &0.549 &3.782 &--0.02 &--3.08 &3.147 \\
CG~Cas &Berkeley~58 &3 &0.6400 &$-3.06\pm0.12$ &0.580 &3.775 &--0.02 &--3.08 &3.149 \\
SZ~Tau &NGC~1647 &4  &0.6512 &$-3.10\pm0.04$ &0.563 &3.779 &--0.02 &--3.12 &3.164 \\
V1334~Cyg &Dolidze~45 &4  &0.6776 &$-3.67\pm0.15$ &0.421 &3.812 &--0.01 &--3.68 &3.388 \\
CV~Mon &van~den~Bergh 1 &3 &0.7307 &$-3.34\pm0.08$ &0.553 &3.781 &--0.02 &--3.36 &3.259 \\
V~Cen &NGC~5662 &5 &0.7399 &$-3.39\pm0.05$ &0.594 &3.772 &--0.02 &--3.41 &3.281 \\
CS~Vel &Ruprecht~79 &... &0.7712 &$-3.56\pm0.22$ &0.506 &3.792 &--0.01 &--3.57 &3.346 \\
V367~Sct &NGC~6649 &... &0.7989 &$-3.51\pm0.07$ &0.556 &3.780 &--0.02 &--3.53 &3.328 \\
BB~Sgr &Collinder~394 &5 &0.8220 &$-3.47\pm0.12$ &0.686 &3.751 &--0.04 &--3.51 &3.320 \\
U~Sgr &IC~4725 &3 &0.8290 &$-3.71\pm0.13$ &0.597 &3.771 &--0.02 &--3.73 &3.410 \\
DL~Cas &NGC~129 &2 &0.9031 &$-3.77\pm0.05$ &0.680 &3.752 &--0.04 &--3.81 &3.440 \\
S~Nor &NGC~6087 &... &0.9892 &$-3.94\pm0.04$ &0.774 &3.732 &--0.07 &--4.01 &3.519 \\
$\zeta$~Gem &ADS~5742 &2 &1.0065 &$-4.10\pm0.10$ &0.756 &3.736 &--0.06 &--4.16 &3.580 \\
TW~Nor &Lyng\aa~6 &... &1.0330 &$-3.99\pm0.18$ &0.645 &3.760 &--0.03 &--4.02 &3.525 \\
V340~Nor &NGC~6067 &... &1.0526 &$-3.80\pm0.17$ &0.866 &3.712 &--0.10 &--3.90 &3.477 \\
X~Cyg &Ruprecht~175 &3 &1.2145 &$-4.88\pm0.07$ &0.886 &3.708 &--0.11 &--4.99 &3.913 \\
RU~Sct &Trumpler~35 &3 &1.2945 &$-4.74\pm0.13$ &0.722 &3.743 &--0.05 &--4.79 &3.832 \\
WZ~Sgr &Turner~2 &3 &1.3394 &$-5.06\pm0.10$ &0.841 &3.717 &--0.09 &--5.15 &3.977 \\
SW~Vel &Vel~OB5 &... &1.3700 &$-5.09\pm0.07$ &0.802 &3.726 &--0.08 &--5.17 &3.983 \\
SV~Vul &Vul~OB1 &2 &1.6540 &$-6.13\pm0.15$ &0.872 &3.711 &--0.11 &--6.24 &4.411 \\
S~Vul &Anon~Vul~OB &3 &1.8325 &$-6.76\pm0.22$ &0.869 &3.712 &--0.10 &--6.86 &4.662 \\
\noalign{\smallskip} \hline
\end{tabular*}
\end{table*}

Previous studies by this author and others have established space reddenings and distances for the clusters and stellar groups associated with several Galactic Cepheids, as well as for the variables themselves. The information available for each calibrating Cepheid was collected and re-examined for this study, and augmented in uncertain cases by new studies of the calibrating clusters and groups using 2MASS photometry \citep{cu03}. The technique of using {\it JHK} magnitudes and colors to deduce reddenings and distances to stellar groups has been illustrated by \citet{te08,te09}. 2MASS photometry is of uneven quality and low precision for most stars in the Galactic plane, but the technique works well in many cases.

New, deep studies of several of the Cepheid calibrating groups have been made by \citet{ho03} and \citet{an07} using CCD {\it UBVK} and {\it BVI$_C$JHK$_S$} observations, respectively. Such studies are useful, but often lead to cluster distances and reddenings that conflict with results from other careful studies. Problems originate from various sources, including the tie-in to standard photometric systems, the data reduction techniques employed, the manner in which cluster main-sequence fitting is used to deduce distances, and, most importantly, the treatment of interstellar reddening as well as differential reddening. The reddening law along the Galactic plane is recognized to vary significantly from one Galactic line of sight to another \citep{wa61,wa62,tu76,tu89}, frequently with the value of $R=A_V/E_{B-V}$ for the dust responsible for extinction correlating closely with reddening slope \citep{tu89,tu94,tu96}. The most effective analyses of cluster distances and reddenings are therefore tied directly to careful examination of the reddening laws for the cluster fields. Accurate determinations of reddening and distance for cluster Cepheids also require the consistent use of reddening parameters that are appropriate for the region being studied \citep{tu76,tu96}.

The list of Cepheids belonging to clusters and groups is presently very extensive, and detailed studies of individual cases are still being undertaken. The number of calibrating Cepheids in well-studied clusters is large enough to provide an independent calibration of the PL relation, to which have been added a number of calibrators resulting from reworkings of existing data or incomplete studies in progress. An example is the reddening of SV Vul, for which 2MASS observations indicate a field reddening of $E_{B-V}= 0.59\pm0.03$ rather than $E_{B-V}= 0.45\pm0.01$ as obtained by \citet{tu84}, although the result is consistent with the OB reddening of $E_{B-V}= 0.65$ obtained by Turner for nearby HDE 339062. In all other cases the cited reddenings were generally confirmed.

\begin{figure}[t]
\begin{center}
\includegraphics[width=0.45\textwidth]{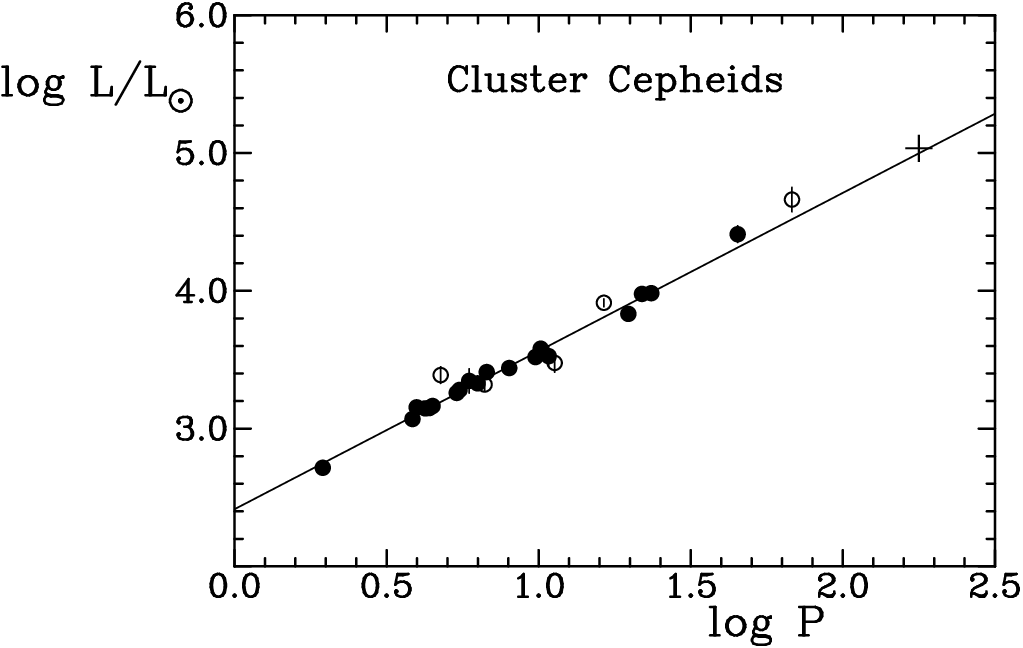}
\end{center}
\caption{\small{The PL relation for Cepheids belonging to open clusters, groups, and associations, with filled circles indicating objects lying near the center of the instability strip and open circles objects near the strip edges. Uncertainties are indicated. A plus sign refers to the Cepheid-like G0 Ia supergiant HD 18391.}}
\label{fig3}
\end{figure}

Revised distance moduli were obtained for several groups.  The distance modulus of Collinder 394, for example, is $9.38\pm0.10$ from 2MASS data, larger than the value of $9.04\pm0.08$ obtained by \citet{tp85} from an incomplete zero-age main-sequence fit. In the case of NGC 5662, the 2MASS data imply an intrinsic distance modulus of $9.28\pm0.05$, larger than the preliminary value of $9.10\pm0.08$ cited by \citet{tu82}, but smaller than the value of $9.50\pm0.15$ obtained by \citet{cl91}. A new study of the cluster is in progress. In the case of Trumpler 35, the 2MASS data imply an intrinsic distance modulus of $11.11\pm0.05$, smaller than the value of $11.58\pm0.18$ obtained by \citet{tu80b} using {\it UBV} data from \citet{ho61} and \citet{yi66}, but consistent with the value of $11.04\pm0.08$ obtained by Yilmaz, who used the Hoag et al. observations for data calibration in an independent {\it UBV} study. In addition, the field of the long-period Cepheid S Vul was found to contain a few B-type stars lying at distances of $\sim$3 kpc that could be used as calibrators. By coincidence, the Cepheid is also projected on a less distant, foreground cluster of late-type stars \citep{te86}. 

The updated results for all cluster Cepheids are summarized for reference purposes in Table~\ref{tab2}, and the resulting data for the Cepheid calibrators are presented in Table~\ref{tab3}, where the cited uncertainties in $\langle M_V \rangle$ represent quadratic sums of the uncertainties in extinction $A_V$ and distance modulus. The Cepheids SZ Tau and V1334 Cyg were considered here to be overtone pulsators, all other Cepheids as fundamental mode pulsators, with the corresponding periods for fundamental mode pulsation derived using the relationship of \citet{sz88}. The magnitudes for $\alpha$ UMi and SU Cyg were corrected for contamination by their unseen companions, but no adjustments were made for other calibrators with likely companions, e.g. X Cyg. Included in Table~\ref{tab3} is an estimate of the instability strip crossing mode for each Cepheid, as deduced from information on rate of period change and light amplitude \citep{te06a}. Use was also made of a current compilation of Cepheid mean magnitudes maintained by \citet{be08}.

The results are plotted in Fig.~\ref{fig3}. The conversion of absolute visual magnitudes and intrinsic ($\langle B \rangle$--$\langle V \rangle$)$_0$ colors into effective temperatures, bolometric magnitudes, and luminosities was done in the manner described by \citet{tb02}. The resulting data provide a tight fit to a unique PL relation that, by coincidence, also fits results for the Cepheid-like G0 Ia supergiant HD 18391 \citep{te09}. The data were analyzed using simple least squares, weighted least squares, and non-parametric techniques, all of which gave similar results, with the average relationship described by:
\begin{displaymath}
\log L/L_{\sun} = 2.418(\pm0.027) + 1.149(\pm0.030) \log P  \;.
\end{displaymath}

Since potential bias could be introduced by Cepheids lying near the edge of the instability strip, the sample was culled to eliminate five Cepheids: V1334 Cyg, BB Sgr, V340 Nor, X Cyg, and S Vul, which have colors that deviate significantly from the instability strip center line (Fig.~\ref{fig1}). The remaining sample of 19 Cepheids was analyzed as previously, although a non-parametric fit was adopted for the final relationship in order to eliminate any remaining potential bias from outliers. The final relationship is relatively unchanged, and is described by:
\begin{displaymath}
\log L/L_{\sun} = 2.415(\pm0.035) + 1.148(\pm0.044) \log P  \;.
\end{displaymath}
The corresponding relationship in terms of absolute magnitude is:
\begin{displaymath}
\langle M_V \rangle = -1.29(\pm0.10) - 2.78(\pm0.12) \log P  \;.
\end{displaymath}

The scatter of the data about the best-fitting relationship in Fig.~\ref{fig3} is small, and in most cases the deviation of individual calibrators from the relationship is linked directly to their location in the instability strip. In only a few cases can the scatter be attributed to poorly established distances and reddenings. The slope of the resulting period-absolute magnitude relation is a good match to previously published estimates based upon open cluster calibrators of $-2.79$ \citep{be85} and $-2.78$ \citep{fw87}, while the zero-point lies between values of $-1.24$ \citep{be85} and $-1.37$ \citep{fw87} obtained previously. Such good agreement is not surprising given that the data for cluster Cepheids used in previous studies have considerable overlap with the sources listed in Table~\ref{tab2}. 

\begin{figure}[t]
\begin{center}
\includegraphics[width=0.45\textwidth]{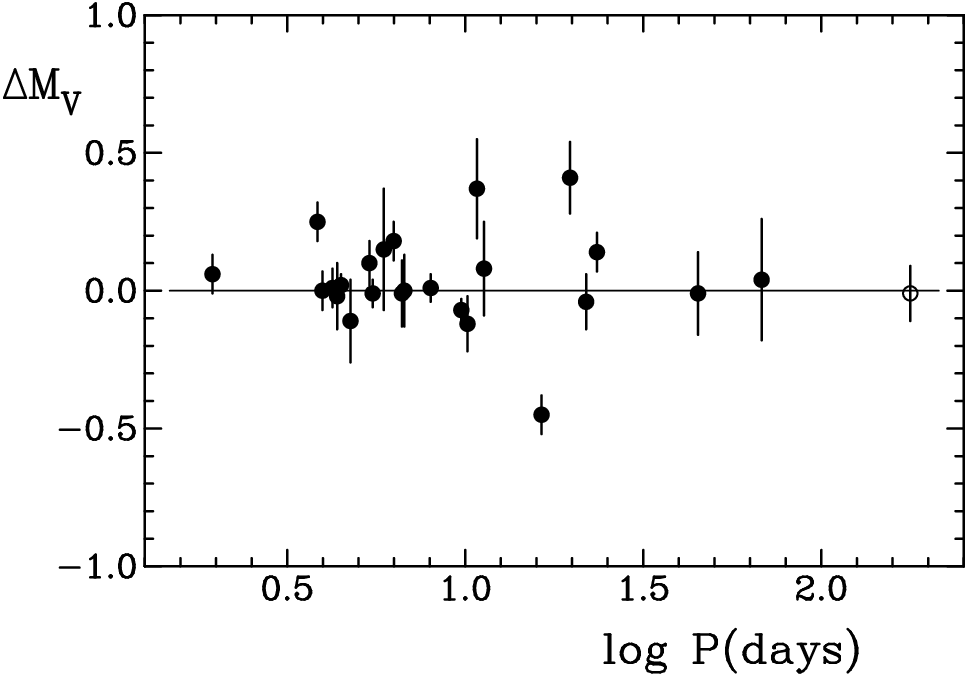}
\end{center}
\caption{\small{Deviations of the calibrating Cepheids from the absolute magnitudes predicted from the period-radius relation in combination with effective temperatures and bolometric corrections tied to intrinsic color, in the sense Observed -- Predicted. The open circle refers to HD 18391.}}
\label{fig4}
\end{figure}

A better comparison is made using absolute magnitudes $\langle M_V \rangle$ inferred from the period-radius relation in conjunction with effective temperatures and bolometric corrections tied to unreddened color. The deviations $\Delta M_V = M_V({\rm observed})-M_V({\rm predicted})$ are plotted in Fig.~\ref{fig4} as a function of pulsation period, with uncertainties as described previously. Here the scatter is also small, with a calculated mean difference of $+0.04\pm0.03$ possibly indicating the need for a slight increment in the cluster distance scale tied to zero-age main-sequence fitting. There is no correlation of the differences with Cepheid metallicity, but that is a particularly challenging test to make with the Table~\ref{tab3} calibrators because of the small spread in metallicities for nearby Cepheids \citep[see][]{tb02}.

\subsection{Parallax Cepheids}

\setcounter{table}{3}
\begin{table*}[t]
\caption[]{Deduced Properties for {\it HST} Parallax Cepheids.}
\label{tab4}
\centering
\small
\begin{tabular*}{0.77\textwidth}{lccccccc}
\noalign{\smallskip} \hline \hline \noalign{\smallskip}
Cepheid &$\log P_0$ &$M_V$ &$(\langle B \rangle - \langle V \rangle)_0$ &$\log T_{\rm eff}$ &B.C. &$M_{\rm bol}$ &$\log L/L_{\sun}$ \\
\noalign{\smallskip} \hline \noalign{\smallskip}
RT~Aur &0.5715 &$-2.88\pm0.18$ &0.520 &3.789 &--0.02 &--2.90 &3.074 \\
T~Vul &0.6469 &$-3.12\pm0.26$ &0.563 &3.779 &--0.02 &--3.14 &3.172 \\
FF~Aql &0.6504 &$-3.11\pm0.14$ &0.531 &3.786 &--0.02 &--3.13 &3.167 \\
$\delta$~Cep &0.7297 &$-3.49\pm0.09$ &0.577 &3.776 &--0.02 &--3.51 &3.321 \\
Y~Sgr &0.7614 &$-3.27\pm0.30$ &0.653 &3.758 &--0.03 &--3.30 &3.238 \\
X~Sgr &0.8459 &$-3.77\pm0.13$ &0.526 &3.787 &--0.02 &--3.79 &3.430 \\
W~Sgr &0.6790 &$-3.93\pm0.19$ &0.625 &3.765 &--0.03 &--3.96 &3.499 \\
$\beta$~Dor &0.9931 &$-3.91\pm0.11$ &0.756 &3.736 &--0.06 &--3.97 &3.504 \\
$\zeta$~Gem &1.0065 &$-3.99\pm0.14$ &0.782 &3.730 &--0.07 &--4.06 &3.540 \\
$\ell$~Car &1.5509 &$-5.29\pm0.22$ &1.095 &3.669 &--0.24 &--5.53 &4.128 \\
\noalign{\smallskip} \hline
\end{tabular*}
\end{table*}

\begin{figure}[t]
\begin{center}
\includegraphics[width=0.45\textwidth]{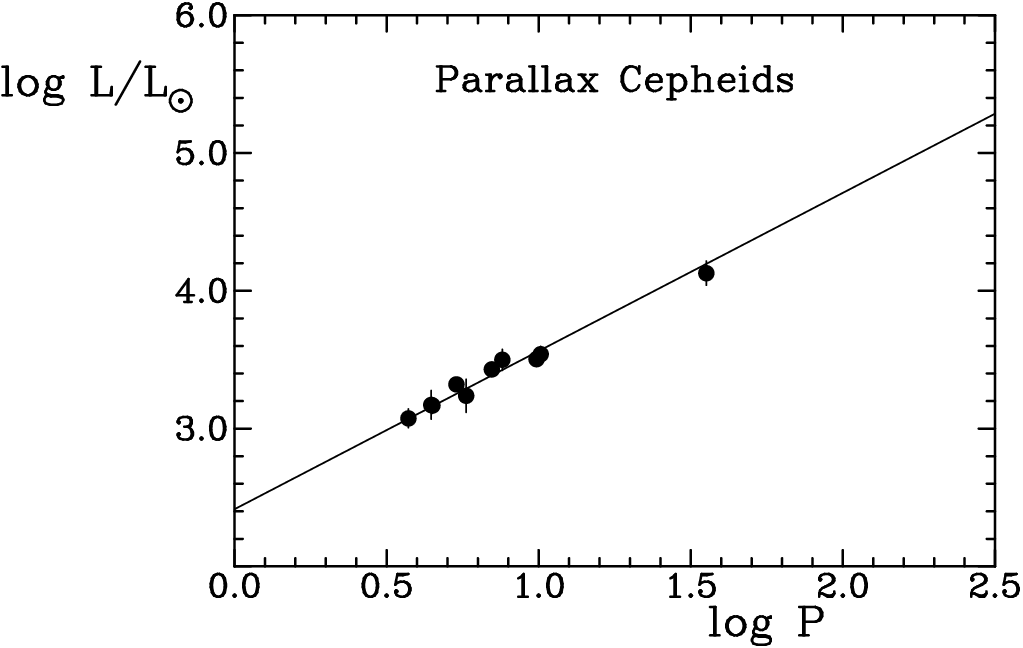}
\end{center}
\caption{\small{The Cepheid PL relation from {\it HST} parallaxes. The plotted relation is that of Fig.~\ref{fig3}.}}
\label{fig5}
\end{figure}

In the case of Cepheids with measured trigonometric parallaxes, it is necessary to have reliable reddenings to correct the magnitudes and colors for the effects of intervening interstellar extinction. In general, such information is available from the studies cited earlier \citep{te87,tu01,bn02,bn07,lc07,ko08}. Table~\ref{tab4} summarizes the available data for {\it HST} Cepheids.

The resulting data for {\it HST} Cepheids are plotted in Fig.~\ref{fig5}, and agree well with the results of Fig.~\ref{fig3} derived for cluster Cepheids. Individual {\it HST} Cepheids deviate from the relationship of Fig.~\ref{fig3} as expected from their location in the instability strip, i.e. X Sgr falls above the relationship because it lies on the hot side of the strip, whereas $\ell$ Car falls below the relationship because it lies on the cool side of the strip.

\begin{figure}[t]
\begin{center}
\includegraphics[width=0.45\textwidth]{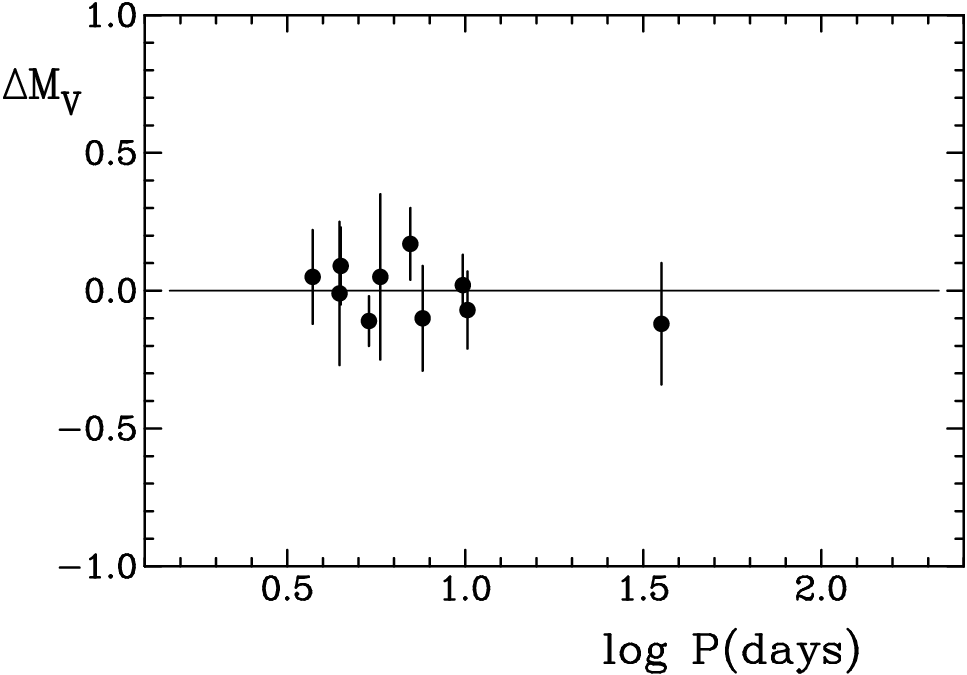}
\end{center}
\caption{\small{Deviations of the {\it HST} parallax Cepheids from the absolute magnitudes predicted from the period-radius relation in combination with effective temperatures and bolometric corrections tied to intrinsic color, in the sense Observed -- Predicted.}}
\label{fig6}
\end{figure}

\begin{figure}[t]
\begin{center}
\includegraphics[width=0.45\textwidth]{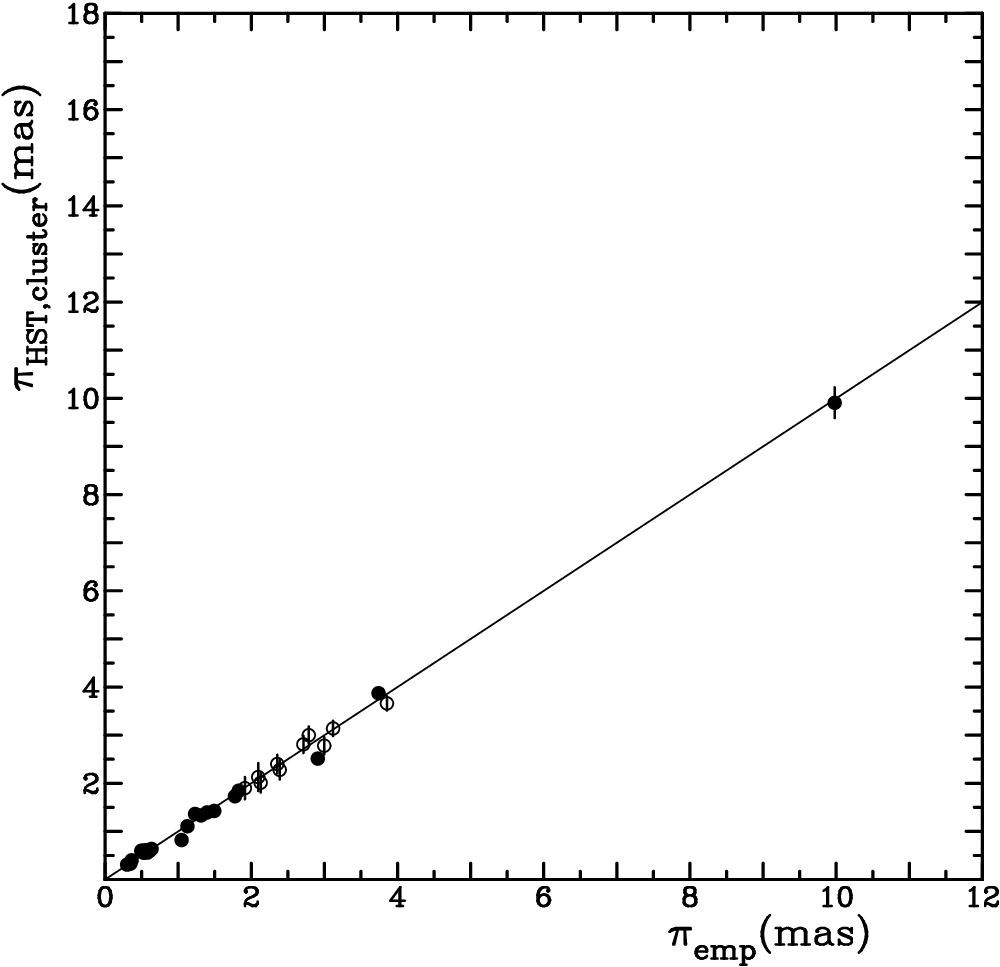}
\end{center}
\caption{\small{A comparison of parallaxes derived for cluster and {\it HST} Cepheids with values predicted from the period-radius relation in combination with effective temperatures and bolometric corrections tied to intrinsic color. Open circles denote {\it HST} Cepheids, filled circles cluster calibrators, and the plotted line is the expected relationship for perfect agreement of the two estimates.}}
\label{fig7}
\end{figure}

As done for cluster Cepheids, the deviations $\Delta M_V = M_V({\rm observed})-M_V({\rm predicted})$ are plotted in Fig.~\ref{fig6} as a function of pulsation period, with the error bars again representing quadratic sums of the uncertainties in extinction and distance modulus. The mean deviation for the {\it HST} Cepheid parallaxes is $+0.00\pm0.03$, which implies that the data confirm the distance scale tied to the period-radius relation used in conjunction with effective temperatures and bolometric corrections obtained from stellar atmospheres. It should be noted, however, that no Lutz-Kelker corrections were applied to the parallaxes in this case. The rationale behind such luminosity adjustments is somewhat unclear \citep{sm03}, despite arguments otherwise \citep{bn07}. The adjustments are typically small, and in the present instance would shift all of the data in Fig.~\ref{fig6} below the null relationship.

An alternate method of comparing implied luminosities for calibrating Cepheids is by plotting their derived parallaxes with those inferred empirically, as presented in Fig.~\ref{fig7}. The small scatter in the data of Fig.~\ref{fig7} emphasizes the good agreement between cluster and {\it HST} parallaxes with predictions from empirical relations. {\it HST} parallaxes for Cepheid calibrators agree with empirical predictions within the stated uncertainties of the former in 7 of 10 cases, or $70\%$, with all three outliers lying within $1.3\sigma$ of the predicted values. In the case of cluster parallaxes for Cepheid calibrators, 16 of 24, or $67\%$, agree with empirical predictions within the stated uncertainties, with all of the outliers lying within $\sim 3\sigma$ of the predicted values except for X Cyg. The observed percentages of calibrators lying within $1\sigma$ of the predicted values are consistent with statistical expectations for both {\it HST} and cluster Cepheids.  

An identical analysis was attempted for the revised {\it Hipparcos} parallaxes of Cepheids tabulated by \citet{vl07}, but with mixed results. An initial sample included only objects in Table~2 of \citet{vl07}, and generated similar results to that obtained for {\it HST} Cepheids, namely $\Delta M_V = M_V({\rm observed})-M_V({\rm predicted}) = +0.02 \pm0.04$, but with larger scatter and the implicit assumption that SU Cas, DT Cyg, BG Cru, and $\alpha$ UMi must be identified as overtone pulsators to be consistent with their observed parallaxes. While the \citet{vl07} sample represents the most suitable selection of {\it Hipparcos} Cepheids for calibration purposes, the question of proper identification of pulsation mode in individual variables makes comparison with the present calibration of uncertain value. 

The identification of pulsation mode in specific Cepheids frequently involves Fourier parameters for their light curves \citep[e.g.,][]{sl81}, as in the case of SU Cas, which is argued to be an overtone pulsator partly on that basis \citep{ai87}. The Fourier parameters for SU Cas are ambiguous regarding its pulsation mode, however, and fundamental mode pulsation is argued by its potential membership in Cas R2 \citep{te84}. Fourier parameters may also not be applicable to sinusoidal Cepheids, the traditional s-Cepheids, which frequently display light curves so closely matching sine waves that their resulting Fourier series do not have significant second and third order terms necessary for generating traditional Fourier parameters. In many such cases the mode of pulsation is not obvious, with fundamental mode and overtone pulsation both being typical of individual variables \citep[see][]{tu09}. A good example is the s-Cepheid FF Aql in Table~\ref{tab1}, a Cepheid that is a fundamental mode Cepheid according to its {\it HST} parallax yet an overtone pulsator according to its {\it Hipparcos} parallax. The higher precision of the {\it HST} parallax would appear to argue the case for the former.

\begin{figure}[t]
\begin{center}
\includegraphics[width=0.45\textwidth]{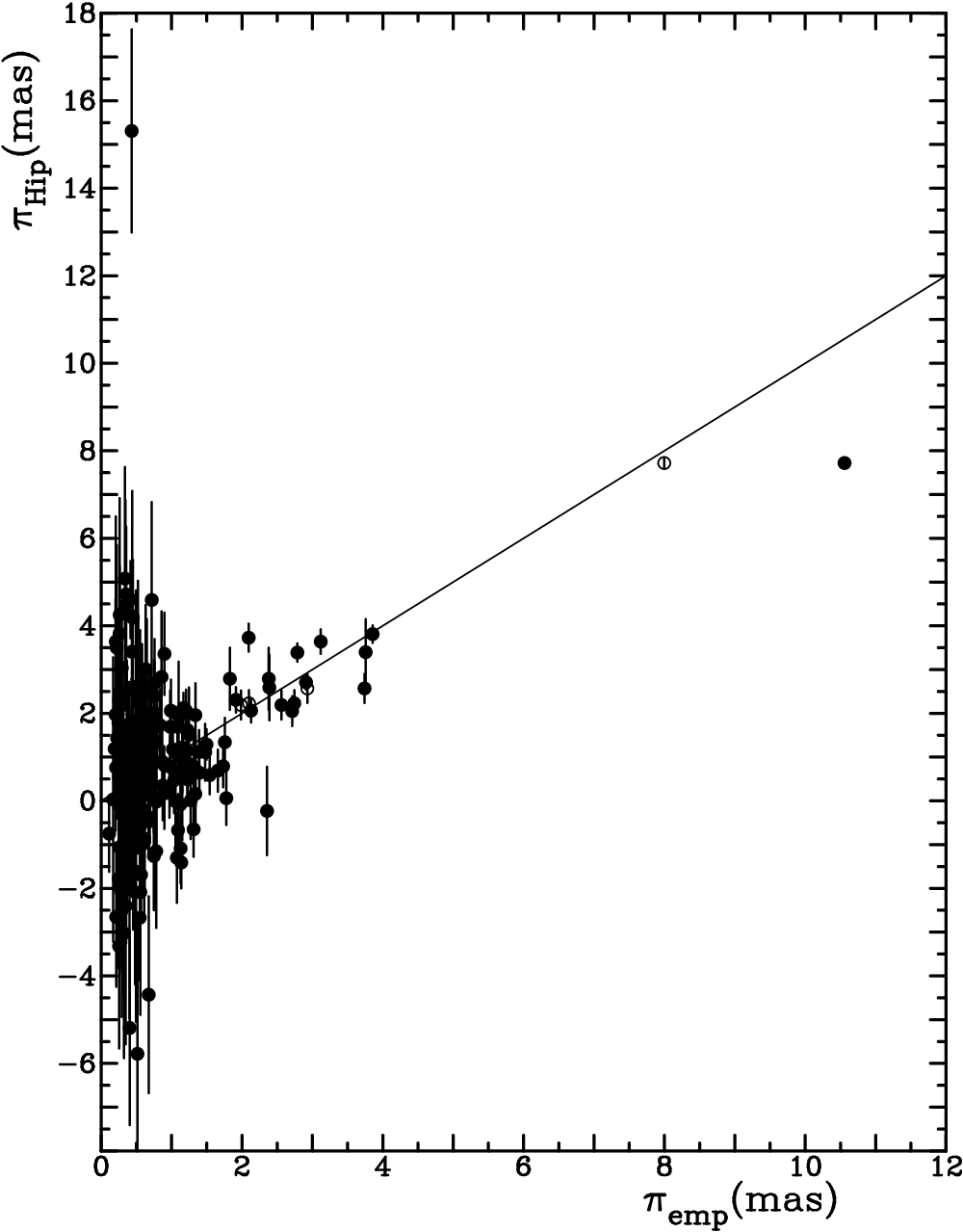}
\end{center}
\caption{\small{A comparison of revised {\it Hipparcos} parallaxes for Cepheids with empirical predictions, in identical fashion to the analysis of Fig.~\ref{fig7}. Open circles denote results for SU Cas, DT Cyg, BG Cru, and $\alpha$ UMi treated as overtone pulsators.}}
\label{fig8}
\end{figure}

It is possible to test how well the Table~2 sample of \citet{vl07} represents the complete set of Cepheids in the {\it Hipparcos} data base. There are 207 such Cepheids with reddenings available from the sources cited above, and all of them were analyzed as above in order to obtain empirical estimates of parallax for comparison purposes. The results are displayed in Fig.~\ref{fig8}, including alternate solutions for four Cepheids suspected to be overtone pulsators. As in Fig.~\ref{fig7}, the relationship represents expectations for exact coincidence of the estimates.

What is evident from Fig.~\ref{fig8} is that natural scatter in the parallaxes is larger for the {\it Hipparcos} Cepheid sample than is the case for {\it HST} and cluster Cepheids. In part that can be attributed to overly-large uncertainties in the parallaxes for faint Cepheids near the observational limits of the {\it Hipparcos} program, as is the case for many of the objects of small parallax in Fig.~\ref{fig8}. But large scatter appears to persist over all parallaxes, even in the case of Cepheids of small $\sigma_{\pi}/\pi$, and is not always resolved by the assumption of overtone pulsation in individual objects. The precision cited for the revised parallaxes of Cepheids in the {\it Hipparcos} collection does not always appear to be representative of the accuracy of the individual parallaxes. Revised {\it Hipparcos} parallaxes for Cepheids agree with empirical predictions within the stated uncertainties of the former in only 108 of 207 cases, or $52\%$ versus expectations closer to $67\%$, although $81\%$ of the outliers (i.e., $91\%$ of the complete sample) fall within $2\sigma$ of the predicted values. Conceivably some parallaxes are affected by systematic offsets, which would make comparisons with individual cluster Cepheids of limited value and could affect conclusions regarding pulsation mode in some objects. As a precaution, {\it Hipparcos} parallaxes were omitted from further steps in the testing procedure.

\section{Discussion and Testing}

\begin{figure}[t]
\begin{center}
\includegraphics[width=0.45\textwidth]{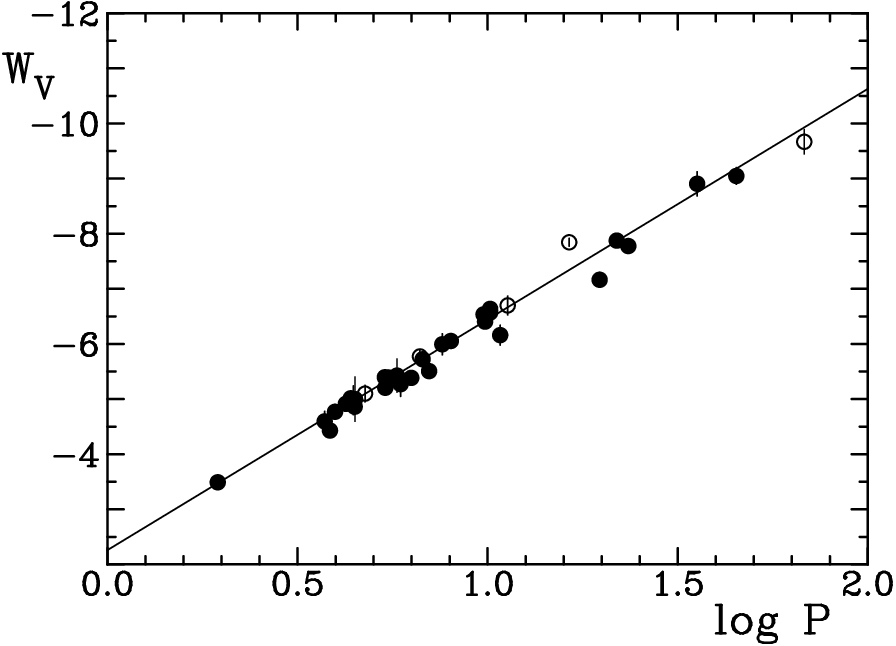}
\end{center}
\caption{\small{The Wesenheit $W_V$ relation from cluster Cepheids and {\it HST} Cepheid parallaxes, with symbols as in Figs. \ref{fig3} and \ref{fig5}.}}
\label{fig9}
\end{figure}

It is important to consider four features of the PL calibration resulting from Cepheids in clusters and groups. First, there exists a very tight linear relationship between the luminosity of a Cepheid and its period of pulsation. Second, the relationship is not ``kinked'' or bent in any fashion. It describes Cepheids of all periods equally well. Third, it is impossible to establish a metallicity dependence for the relationship because of the small spread in [Fe/H] values for existing Galactic calibrators. Lastly, the cluster calibration of the Cepheid PL relation exhibits a scatter amounting to only $\pm0.17$ s.d. for individual calibrators, or $\pm0.03$ s.e. for the entire sample. That may improve once detailed studies are complete for some of the Cepheids in Table~\ref{tab3} for which only preliminary results are available.

A comparison with similarly-derived luminosities for Cepheids with well established {\it HST} parallaxes \citep{bn02,bn07} produces essentially identical results, although the period coverage is less uniform. The unusual period-absolute magnitude relationship resulting from {\it HST} parallaxes \citep{bn07} is demonstrated to be a consequence of a pathological choice of calibrators: the long period Cepheids ($P \ge 9^{\rm d}$) in the sample all lie on the cool side of the instability strip and have overly small luminosities, and the short period Cepheids all lie towards the center or hot side of the instability strip and have overly large luminosities. The selection of Cepheid calibrators for the PL relation clearly affects the results obtained.

An important factor affecting such an analysis is the reddening of individual Cepheids. Given that Wesenheit formulations are generally designed to correct for the effects of interstellar reddening, they would appear to be the instrument of choice for studies of Cepheid distances. There is an alternative available using luminosities directly, as demonstrated here, but it is difficult to apply such a relationship in distant galaxies where only visible and near-infrared photometry is typically available for member Cepheids. A Wesenheit formulation can be derived from the $M_V$ data of Table~\ref{tab3}, adjusted by --0.04 to account for the need of a possible zero-point shift in the calibrated zero-age main sequence, and the {\it HST} parallax data in Table~\ref{tab4}, as shown in Fig.~\ref{fig9}. For this purpose a color term of 3.3 was adopted \citep[see][]{mf91}. A non-parametric fit to the data produces a relationship described by:
\begin{displaymath}
W_V = -2.259(\pm0.083) - 4.185(\pm0.103) \log P  \;.
\end{displaymath}

\begin{figure}[t]
\begin{center}
\includegraphics[width=0.45\textwidth]{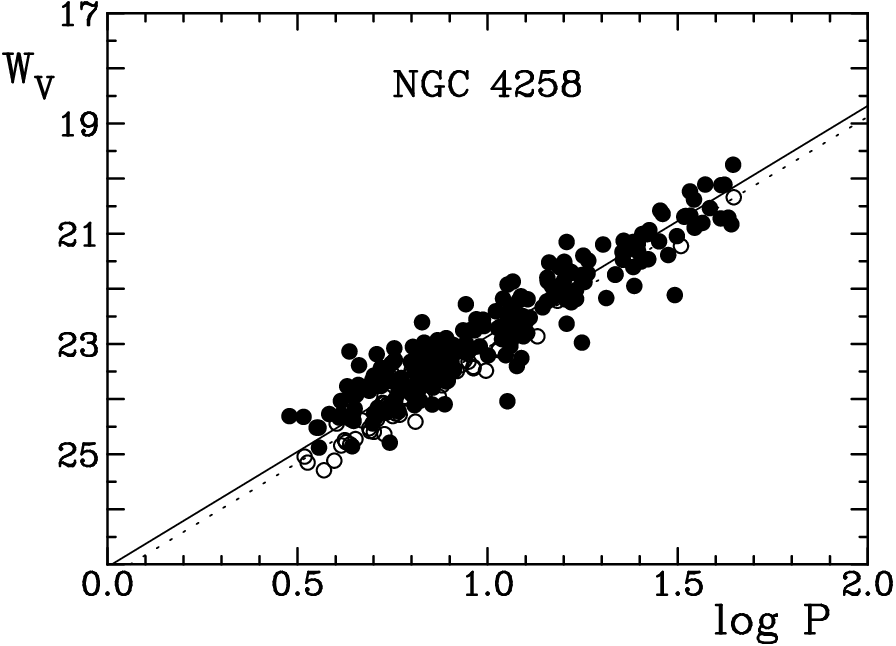}
\end{center}
\caption{\small{The Wesenheit $W_V$ relation for Cepheids in NGC 4258. Filled circles refer to inner region Cepheids, and open circles to outer region Cepheids, with solid and dotted lines representing the solutions for the two groups, respectively.}}
\label{fig10}
\end{figure}

As a test of the validity of the resulting relationship for extragalactic Cepheids, an analysis was made using the data from \citet{mc06} for Cepheids in the inner regions of NGC 4258, where metallicities appear to be close to the solar values typical of nearby Galactic calibrators. The available {\it BV} data for the galaxy's Cepheids \citep{mc06} were analyzed using the Wesenheit formulation, as shown in Fig.~\ref{fig10}. Although the scatter is larger than for Galactic calibrators, the agreement is otherwise excellent, particularly the trend with period, which depends heavily upon the Wesenheit formulation and the validity of the Galactic calibration over all pulsation periods. Restricting the sample to inner region Cepheids with $P \ge 12^{\rm d}$, as suggested by \citet{mc06} as a means of eliminating bias arising from contaminated Cepheids, and eliminating a few outliers yields a distance modulus for NGC 4258 of $29.31\pm0.04$, corresponding to a distance of $7.27\pm0.13$ Mpc. That value agrees closely with the geometrical distance estimate of $7.2\pm1.4$ Mpc obtained by \citet{he99} using OH masers. Cepheids in the outer regions of NGC 4258 yield a distance modulus of $29.51\pm0.05$, the $0^{\rm m}.2$ difference relative to the value obtained for Cepheids lying in the inner regions of NGC 4258 presumably being tied to the lower metallicity of outer region Cepheids.

The sole remaining factor affecting Cepheid distances remains the treatment of interstellar and intergalactic reddening. The zero-point established by Galactic calibrators lying in calibrating clusters and groups or having {\it HST} parallaxes appears to be very solid, and eliminates the need to rely on Magellanic Cloud Cepheids \citep{bn07,fo07} or surface brightness relations \citep{fo07} to establish the extragalactic distance scale. The near-solar metallicities of most Galactic calibrators also make them more similar to the Cepheids sampled in other spiral galaxies, an important factor when adjusting distances for metallicity effects, certainly in the {\it B} and {\it V} bands used here. The manner of treating reddening affects the Wesenheit formulation, however, and may also influence the technique of PL-fitting in different wavelength bands ({\it e.g.}, {\it R} and {\it I}) that are used in most extragalactic studies. Any differences between the extinction properties of dust in nearby regions of the Galaxy and those in other galaxies may yet have an important influence on the extragalactic distance scale.

\end{document}